\documentclass[%
 reprint,
 amsmath,amssymb,
 aps,
]{revtex4-2}

\usepackage{graphicx}
\usepackage{dcolumn}
\usepackage{bm}

\begin{document}


\title{Higher-order NLO radiative corrections to polarized muon decay spectrum }

\author{A.~B.~Arbuzov}
\email{arbuzov@theor.jinr.ru}
\author{U.~E.~Voznaya}%
 \email{voznaya@theor.jinr.ru}
 \affiliation{Joint Institute for Nuclear Research,             
 Joliot-Curie str. 6, Dubna, 141980, Moscow region, Russia \\
 Dubna state university, Universitetskaya str. 19, Dubna, 141980, 
 Moscow region, Russia}

\date{\today}

\begin{abstract}
Higher-order QED radiative corrections to muon decay spectrum are evaluated within
the QED structure function approach in the next-to-leading order logarithmic approximation.
New analytical results are given in the 
$\mathcal{O}\left(\alpha^3 \ln^2(m_\mu^2/m_e^2)\right)$ order. Earlier results in 
$\mathcal{O}\left(\alpha^2 \ln^1(m_\mu^2/m_e^2)\right)$ and 
$\mathcal{O}\left(\alpha^3 \ln^3(m_\mu^2/m_e^2)\right)$ orders are partially corrected.
Numerical estimates of different contributions are presented.
\end{abstract}

\maketitle


\section{\label{sec:level1}Introduction}

Studies of muon decay 
\begin{equation}
\mu^- \longrightarrow e^- + {\bar{\nu}}_e +\nu_{\mu}
\end{equation}
are one of the cornerstones of modern particle physics.
This process is almost a pure weak-interaction process with small QED, QCD, and possibly new physics additions. 
High-precision and high-sensitivity experiments with muons can test small deviations from 
the Standard Model (SM) predictions, which would be the traces of new physics. 
Differential distributions in muon decays allows studying properties of weak interactions, 
including even the Dirac or Majorana nature of neutrinos~\cite{Szafron:2009zz,Marquez:2022bpg}. 
Such experiments as TWIST \cite{Mischke:2008ed, Bayes:2013esa}, Mu2e \cite{Bernstein:2019fyh}, Mu3e \cite{Hesketh:2022wgw} 
require accurate advanced theoretical predictions. Precision of the predictions can be increased 
by calculation of higher-order radiative corrections. 

QED corrections to the muon lifetime are known from the works~\cite{Behrends:1955mb, Berman:1962uvx, Kinoshita:1958ru, Berman:1958ti, Steinhauser:1999bx, vanRitbergen:1998yd, Czarnecki:2001cz} upto the  $\mathcal{O}(\alpha^2)$ order,
and the $\mathcal{O}(\alpha^3)$ corrections were also recently calculated~\cite{Fael:2020tow}. 
The TWIST experiment required corrections 
to the muon decay spectrum in at least the $\mathcal{O}(\alpha^2)$ order. 
In ref.~\cite{Arbuzov:2002cn} radiative corrections to unpolarized 
muon decay spectrum to the order $\mathcal{O}(\alpha^2 L$) where $L\equiv\ln(m_\mu^2/m_e^2)$, 
and in ref.~\cite{Arbuzov:2002rp} radiative corrections to the polarized muon decay 
spectrum in the $\mathcal{O}(\alpha^3 L^3)$ and $\mathcal{O}(\alpha^2 L)$ orders were calculated analytically.  
Complete two-loop QED corrections to muon decay spectrum were calculated numerically~\cite{Anastasiou:2005pn} 
in a restricted kinematics domain.
Recently, the results for the pure-photonic part of the higher-order QED corrections in the leading 
and next-to-leading logarithmic approximations were presented in~\cite{Banerjee:2022nbr}.

Our aim here is to calculate $\mathcal{O}(\alpha^3 L^2)$ order corrections to the electron energy spectrum in 
decays of polarized and unpolarized muons. 
The paper is organized as follows. In the next Sect.
we describe application of the QED structure function formalism for calculation of corrections to the muon decay spectrum. Sect.~\ref{sect:num} contains numerical results and a discussion about the factorization scale choice. Lengthy formulae with analytic results are shifted to Appendices. 

\section{Corrections to electron energy spectrum}

Analytic calculations of higher-order radiative corrections
to muon decay spectrum as well as to differential distributions
of other processes like Bhabha scattering and electron-positron annihilation 
are very difficult because of the presence of several energy scales. 
Only a few complete results for $\mathcal{O}(\alpha^2)$ QED
radiative corrections to differential distributions are known.

On the other hand, the bulk of QED radiative corrections typically comes
from the terms enhanced by powers of the large logarithms
$\ln Q^2/m^2$, where $Q^2$ is the square of the characteristic energy scale 
and $m$ is the mass of a light charged lepton, e.g., 
$L=\ln M_Z^2/m_e^2\approx 24$ for the process of $e^+e^-$ annihilation into $Z$ boson.

In the QED structure function approach~\cite{Kuraev:1985hb}, one can get corrections enhanced by the large logarithms
by performing a convolution of a hard scattering cross section and the corresponding parton distribution or fragmentation 
functions which are independent of the process. In general, the large logarithm reads
\begin{equation}
L = \ln \frac{\mu_F^2}{\mu_R^2},
\end{equation}
where $\mu_F$ is the factorization scale and $\mu_R$ is the renormalization scale. 
If $\mu_F\gg\mu_R$, the corrections proportional to powers of $L$ yield the most significant contributions. 
In the case of muon decay, we can calculate the electron energy spectrum in the following way~\cite{Arbuzov:2002cn}:
\begin{eqnarray}
\frac{d \Gamma}{d c d x} = \sum \limits_{j=e, \gamma} \int \limits_x^1 \frac{d z}{z} \frac{d^2 \hat{\Gamma}_j }{dcd z} (z, c, \mu_F, \mu_R) 
 D_{ej} \left( \frac{x}{z}, \mu_F, \mu_R \right),
\end{eqnarray}
where $c$ is the cosine of the angle $\theta$ between muon polarization vector and electron momentum. 
Above, $z$ is the energy fraction of the parton $j$ produced in muon decay, $x$ is the energy fraction of 
the resulting massive electron, $d \hat{\Gamma}_j /(dc dz)$ is the energy and angle distribution of 
the massless parton $j$, $D_{ej}$ is the fragmentation function that describes the probability density for
transformation of the massless parton $j$ into the physical electron in the final state, 
$\mu_F$ is the factorization scale. Here the standard $\mathrm{MS}$ subtraction scheme is used.
We can take $\mu_F = m_\mu$ and $\mu_R = m_e$, so the large logarithm is
\begin{equation}
L = \ln \frac{m_{\mu}^2}{m_e^2} \approx 10.66.
\end{equation}

The perturbative expansion of the kernel coefficient function $d \hat{\Gamma}_j /(dc dz)$ in powers of $\alpha$ reads
\begin{eqnarray}
	\frac{1}{\Gamma_0} \frac{d \hat{\Gamma}_j }{dc d z} (z, \mu_F,\mu_R) = A^{(0)}_j (z) + \frac{\alpha (\mu_F^2)}{2 \pi} \hat{A}_j^{(1)}(z) \nonumber \\
 + \left( \frac{\alpha (\mu_F^2)}{2 \pi}  \right) \hat{A}_j^{(2)}(z) + ...
\end{eqnarray}
where $\Gamma_0 = G^2_F m^5_{\mu} /(192 \pi^3 )$, $A^{(0)}_j (z) = z^2 (3-2z) \delta_{je}$, $\alpha (\mu_F^2)$ 
is the renormalized fine structure constant taken at the factorization energy scale.

Here we used process-independent fragmentation functions, calculated by solving the QED evolution equation
\begin{eqnarray} \label{evol_eq}
&&	D_{ba}(x, \frac{\mu^2_R}{\mu^2_F}) = \delta(1-x)\delta_{ba} + \nonumber \\
&&  \sum\limits_{i=e,\bar{e},\gamma}\int\limits_{\mu_R^2}^{\mu_F^2} 
	\frac{dt \alpha(t)}{2 \pi t}  \int\limits_{x}^{1} \frac{dy}{y} D_{ia}(y, \frac{\mu^2_R}{t}) P_{bi} \left( \frac{x}{y}, t \right).
\end{eqnarray}
For the initial conditions and other details, see ref.~\cite{Arbuzov:2022fmv}.
Note that here and in what follows we apply the natural choice of the QED renormalization constant 
$\mu_R = m_e$. 

In the unpolarized case, the relevant coefficient functions are
\begin{equation}
\hat{A}^{(0)}_{U,e} (z)= f_e^{(0)} (z), \qquad \hat{A}^{(1)}_{U,e} (z) = f_e^{(1)} (z),
\end{equation}
\begin{equation}
\hat{A}^{(0)}_{U,\gamma} (z) = 0, \qquad \hat{A}^{(1)}_{U, \gamma} (z)= f_\gamma^{(1)} (z).
\end{equation}
In the polarized case, the corrections include the parts which are dependent on the polarization 
degree $P_\mu$ and the cosine of the angle between the muon spin and electron's momentum $c = cos \theta$,
\begin{eqnarray}
&& \hat{A}^{(0)}_{P,e} = f_e^{(0)} (z) + c P_\mu g_e^{(0)} (z), \\
&& \hat{A}^{(1)}_{P,e} (z) = f_e^{(1)} (z) + c P_\mu g_e^{(1)} (z), \\
&& \hat{A}^{(0)}_{P,\gamma} (z) = 0, \\
&& \hat{A}^{(1)}_{P, \gamma} (z)= f_\gamma^{(1)} (z) + c P_\mu g_\gamma^{(1)} (z).
\end{eqnarray}

The expression for the differential distribution of electrons (averaged over electron spin states) 
in a polarized muon decay reads \cite{Arbuzov:2002rp}
\begin{eqnarray}
&& \frac{d^2 \Gamma}{d z d c} = \Gamma_0 (F(z) \pm  c P_\mu G(z)), \quad 
z = \frac{2 m_\mu E_e}{m^2_\mu + m_e^2}, 
\nonumber \\ && 
 z_0 \leq z \leq 1, \quad z_0 = \frac{2 m_\mu m_e}{m^2_\mu + m_e^2},
\end{eqnarray}
where "+" and "-" correspond to $e^+$ and $e^-$ respectively, $G_F$ is Fermi coupling constant, 
$E_e$ and $z=2E_e/m_\mu$ are the energy and the energy fraction of the electron (or positron), respectively.

The complete expression for the spectrum functions ($H=F,G$) up to the $\mathcal{O}({\alpha^3L^2})$ order reads
\begin{eqnarray} \label{eq:F_z}
 && H(z) = h_e^{(0)}(z) + \frac{\alpha}{2 \pi} h_1  + \left( \frac{\alpha}{2 \pi} \right)^2 \Bigl\{ \Bigl[ h_2^{(0, \gamma)} + h_2^{(0, NS)} \nonumber \\ 
 && + h_2^{(0, S)} \Bigr] \frac{L^2}{2}+ \Bigl[ h_2^{(1, \gamma)} +  h_2^{(1, NS)} +h_2^{(1, S)} + h_2^{(1, int)} \Bigr] L\Bigr\}  \nonumber \\  
 && + \left( \frac{\alpha}{2 \pi} \right)^3 \Bigl\{ \Bigl[ h_3^{(0, \gamma)} + h_3^{(0, NS)} + h_3^{(0, S)} \Bigr]\frac{L^3}{6} + \Bigl[ h_3^{(1, \gamma)} 
  \nonumber  \\  
 && + h_3^{(1, NS)} +h_3^{(1, S)} + h_3^{(1, int)} \Bigr]\frac{L^2}{2}  \Bigr\}  \nonumber  \\  
 && \equiv h_e^{(0)}(z)  + \sum_{i,j}  \alpha^i L^j H_{ij} (z),
\end{eqnarray}
where indices $\gamma$, $NS$, $S$ and $int$ correspond to the pure photonic contribution, the non-singlet fermion pair one, 
the singlet fermion pair one, and the interference of the singlet and non-singlet pair corrections. Here and in what follows we omit the arguments of the functions $h_i$ for convenience with $h\equiv f$ or $h\equiv g$.

To get the contribution of the order $\mathcal{O}(\alpha^3 L^2)$, we have to make convolutions of 
the fragmentation functions with functions $h^i_{e}(z)$ and $h^i_{\gamma}(z)$:
\begin{eqnarray}
 \label{F32conv}
    && \left(h_e^{(0)}(z) + \frac{\alpha}{2 \pi}h_e^{(1)} (z)\right) \otimes \bigl[ D_{ee} \bigr]_T 
  \nonumber  \\ && \quad
    + \left(h_\gamma^{(0)} (z)+ \frac{\alpha}{2 \pi}h_\gamma^{(1)} (z) \right) \otimes \bigl[ D_{e \gamma} \bigr]_T,
\end{eqnarray}
and take only the terms proportional to $\alpha^3 L^2$ from the result.

Expressions for the polarized part can be received from the equation (\ref{F32conv}) by substitution $f_i^{(j,r)} \rightarrow g_i^{(j,r)}$. Index $T$ in the above equation marks (timelike) fragmentation functions. 


We recalculated the $\mathcal{O}(\alpha^2 L)$ corrections and found that the term 
$d_{\gamma e}^{(1)}(x) \otimes P_{e \gamma}^{(0)} $   (see Eq.~(\ref{DeeS}))
has been missed in the electron fragmentation function used in Ref.~\cite{Arbuzov:2002cn} . 
The difference is
\begin{eqnarray}
&& F_{21}^{new} (z)-F_{21}^{old} (z)=\frac{46 z^3}{27}+\frac{151 z^2}{6} -19 z-\frac{8}{3z} 
- \frac{281}{54} 
   \nonumber \\ &&
  - \biggl( \frac{64}{9} + \frac{4}{3 z} + 18 z + \frac{38 z^2}{3}\biggr) \ln z  - \left(  \frac{5}{3} + 4z\right) \ln^2 z.
\end{eqnarray}
A similar change with respect to the result given in Ref.~\cite{Arbuzov:2002rp} is found for function $G_{21}(z)$, 
the corrected expression for it is shown in Appendix~\ref{appendix:Functions}.

In our previous work \cite{Arbuzov:2022fmv} we also corrected a mistake in the result for the 
$\mathcal{O}(\alpha^3 L^3)$ singlet contribution to structure and fragmentation functions
obtained in Ref.~\cite{Skrzypek:1992vk}. As the result, the
singlet part of the NLO electron fragmentation function should read
\begin{eqnarray}
&& \bigl[ D_{ee}^{ S} \bigr]_T = \left(\frac{\alpha}{2 \pi} \right)^2 L \left(P_{ee}^{(1), S} +  d_{\gamma e}^{(1)}(x) \otimes P_{e \gamma}^{(0)} \right)\nonumber \\
&&  + \Bigl( \frac{\alpha}{2 \pi} \Bigr)^2 L^2 
	 \frac{1}{2} P_{\gamma e}^{(0)} \otimes P_{e \gamma}^{(0)} +    \Bigl( \frac{\alpha}{2 \pi} \Bigr)^3 L^3 \biggl( \frac{1}{6}  P_{ee}^{(0)} \otimes P_{ee}^{(0)} 
  \nonumber \\ && 
  \otimes P_{ee}^{(0)}
  + \frac{2}{9} P_{\gamma e}^{(0)} \otimes  P_{e \gamma}^{(0)}\biggr) + \mathcal{O}(\alpha^3L^2).
\end{eqnarray}

Functions $f^{(i)}_a$ and $g^{(i)}_a$ read~\cite{Kinoshita:1958ru}:
\begin{eqnarray}
&& f_e^{(0)} (z)= z^2 (3 - 2z), \\
&& f_e^{(1)} (z) = 2 z^2 (2 z-3) \bigl( 4 \zeta(2) - 4 \mathrm{Li}_2 (z) + 2  \ln z^2 
        \nonumber
\\ 
&& - 3  \ln z  \ln (1-z)  -  \ln (1-z)^2 \bigr) + \biggl( \frac{5}{3} - 2 z - 13 z^2 
     \nonumber \\
&&  + \frac{34}{3} z^3 \biggr)  \ln (1-z) + \left( \frac{5}{3} + 4 z - 2 z^2 - 6 z^3 \right)  \ln z  
\nonumber \\ &&  
+ \frac{5}{6}  
- \frac{23}{3} z -\frac{3}{2} z^2 + \frac{7}{3} z^3, \\
&& f_\gamma^{(0)} (z)= 0, \\
&& f_\gamma^{(1)} (z)= \ln z   \left(  - \frac{10}{3} + \frac{2}{z} + 4 z \right)
      + \ln (1-z)   \biggl(  - \frac{5}{3} + \frac{1}{z} \nonumber \\
&& + 2 z - 2 z^2 + \frac{2}{3} z^3 \biggr)
+ \frac{1}{3} - \frac{1}{z} + \frac{35}{12} z - 2 z^2 - \frac{1}{4} z^3,\\
&& g_e^{(0)} (z) = z^2 (1-2 z) , \\
&& g_e^{(1)} (z) =  2 z^2 (1-2 z) \biggl(\ln (1-z)^2 - 4 \mathrm{Li}_2 (1-z) \nonumber \\
&&- \ln (z) \ln (1-z) - 2 \ln (z)^2 \biggr) + \biggl( \frac{11}{3} - \frac{4}{3 z}  - 6 z \nonumber \\
&&
 - \frac{17}{3} z^2 + \frac{34}{3} z^3\biggr) \ln (1-z) + \left(-\frac{1}{3} - 6 z^2 - 6 z^3 \right) \ln (z) \nonumber \\
&&  - \frac{7}{6} + 3 z + \frac{7}{6} z^2 + 3 z^3, \\
&& g_\gamma^{(0)} (z)= 0, \\
&& g_\gamma^{(1)} (z)= \left(\frac{1}{3} - \frac{1}{3 z} - \frac{2}{3} z^2 + \frac{2}{3} z^3 \right) \ln (1-z)
    \nonumber \\
&& + \biggl(\frac{2}{3}- \frac{2}{3 z}\biggr) \ln z - \frac{2}{3} + \frac{2}{3 z} + \frac{11}{12} z - \frac{2}{3} z^2 - \frac{1}{4} z^3.
\end{eqnarray}

The relevant fragmentation functions are shown in the Appendix~\ref{results}, see ref.~\cite{Arbuzov:2022fmv} for details of notation and explicit expressions for these functions. 

Convolutions were calculated using our own program in FORM~\cite{Vermaseren:2000nd} 
and crosschecked with the help of the HPL~\cite{Maitre:2005uu} and MT~\cite{Hoschele:2013pvt} 
Wolfram Mathematica packages.
The results are presented in Appendix~\ref{results}. A part of the results for the unpolarized case were 
presented in \cite{Voznaya:2023enr}, here we reproduce them for the sake of completeness.

We calculated separately the parts of $F_{21}, F_{22}, F_{43}, F_{44}$ and 
$G_{21}, G_{22}, G_{43}, G_{44}$ with  pure-photonic contributions in order to compare with the results of Ref.~\cite{Banerjee:2022nbr}. 
Our results completely agreed with the ones from this work in the orders 
$\mathcal{O}(\alpha^3 L^2)$, $\mathcal{O}(\alpha^4 L^4)$, and $\mathcal{O}(\alpha^4 L^3)$. 

\section{Factorization scale choice and numerical results} 
\label{sect:num}

\begin{figure}[h!]
 \includegraphics[width=0.9\linewidth]{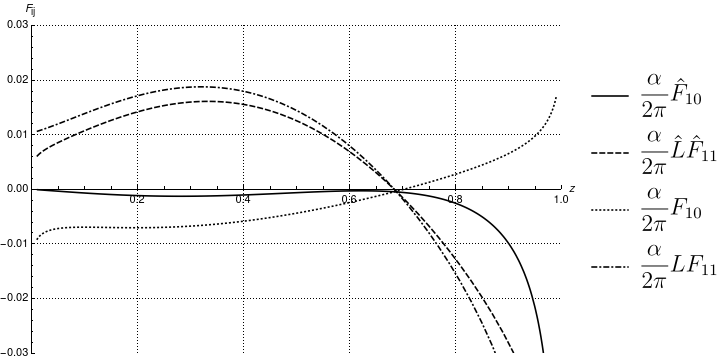}
		\caption{Contributions in $\mathcal{O}(\alpha^1 L^0)$ and $\mathcal{O}(\alpha^1 L^1)$ orders for the old and new factorization scales.} \label{1it}
\end{figure}

\begin{figure}[h!]
 \includegraphics[width=0.9\linewidth]{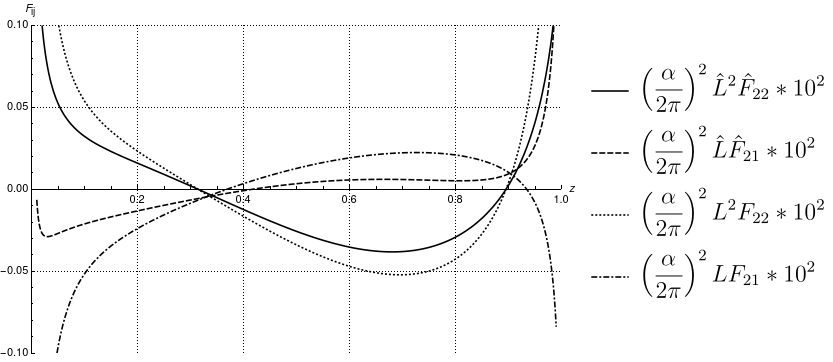}
		\caption{Contributions in $\mathcal{O}(\alpha^2 L^2)$ and $\mathcal{O}(\alpha^2 L)$ orders for the old and new factorization scales.} \label{2it}
\end{figure}

\begin{figure}[h!]
		\includegraphics[width=0.9\linewidth]{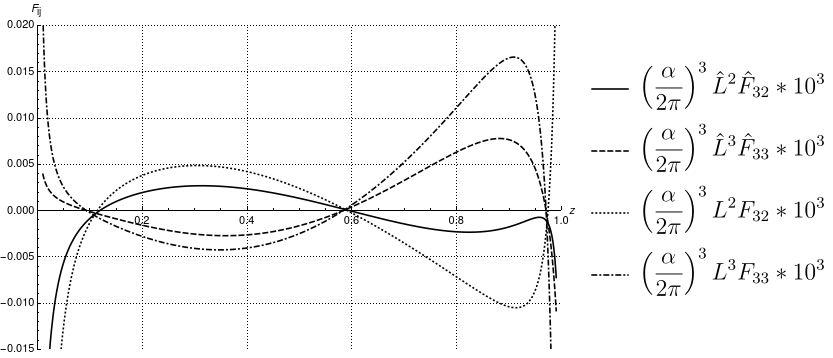}
		\caption{Contributions in $\mathcal{O}(\alpha^3 L^3)$ and $\mathcal{O}(\alpha^3 L^2)$ orders for the old and new factorization scales.} \label{3it}
\end{figure}

The factorization scale choice allows some arbitrariness. Above and in earlier 
papers~\cite{Arbuzov:2002cn, Arbuzov:2002rp} the muon mass was taken as the factorization scale. 
This choice is certainly good for the leading logarithmic approximation, but can be optimized if 
one goes beyond it. We suggest choosing the factorization scale as
\begin{equation}
\mu_F^2 = m_{\mu}^2 z(1-z).
\end{equation}
So we expand on the powers of new large logarithm:
 \begin{equation}
 \hat{L} = L + \Delta L, \quad \Delta L = \ln z + \ln (1-z).
 \end{equation}
With this choice, NLO contributions are shifted by an additional term
\begin{equation}
\hat{F}_{ab} = F_{ab} - 2 \Delta L F_{aa},\qquad b=a-1,
\end{equation}
and the same for the $G$ part. Here indices $a$ and $b$ are powers of $\alpha$ and $L$, respectively; 
$\hat{F}_{ab}$ is the NLO contribution for the new factorization scale choice and $F_{ab}$ is for the old one.
The LO contributions do not change,
\begin{equation}
\hat{F}_{aa} = F_{aa}.
\end{equation}
The new factorization scale choice increases the difference between the NLO and LO contributions, thus improving the
convergence of the expansion in the powers of the large logs. The results for the
two factorization scale choices are shown on the plots for $\mathcal{O}(\alpha^1)$ (Fig.~\ref{1it}), 
$\mathcal{O}(\alpha^2)$ (Fig.~\ref{2it}), and $\mathcal{O}(\alpha^3)$ (Fig.~\ref{3it}).


\begin{figure}[h!]
		\includegraphics[width=0.9\linewidth]{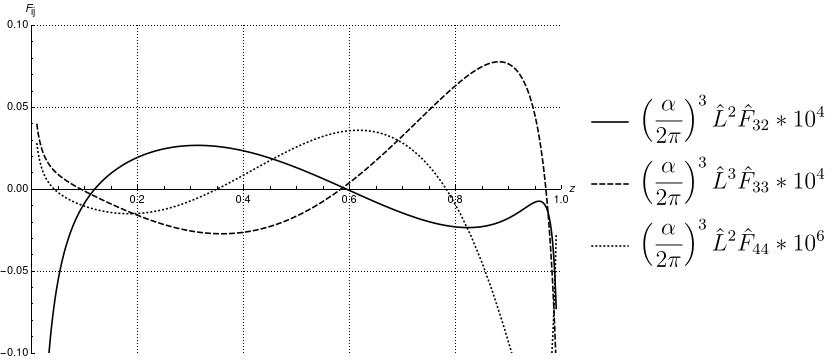}
		\caption{$F$-part of the corrections of the orders 
  $\mathcal{O}(\alpha^3 L^2,\ \alpha^3 L^3,\ \alpha^4 L^4)$ for the new factorization scale.} \label{F3}
\end{figure}
\begin{figure}[h!]
 \includegraphics[width=0.9\linewidth]{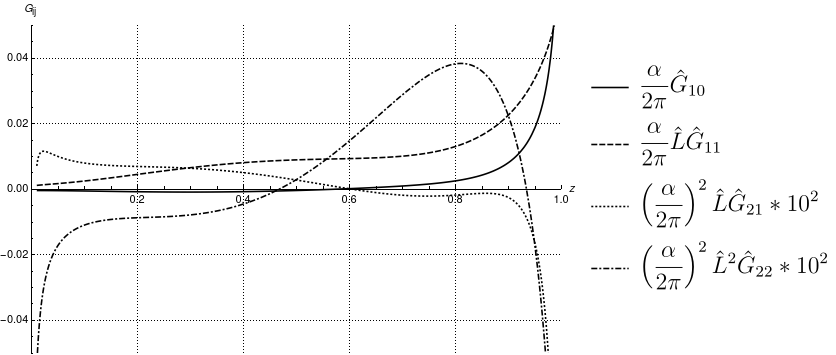}
		\caption{$G$-part of the corrections of the orders $\mathcal{O}(\alpha,\ \alpha L,\ \alpha^2 L,\ \alpha^2 L^2)$ for the new factorization scale.} \label{G1}
\end{figure}
\begin{figure}[h!]
		\includegraphics[width=0.9\linewidth]{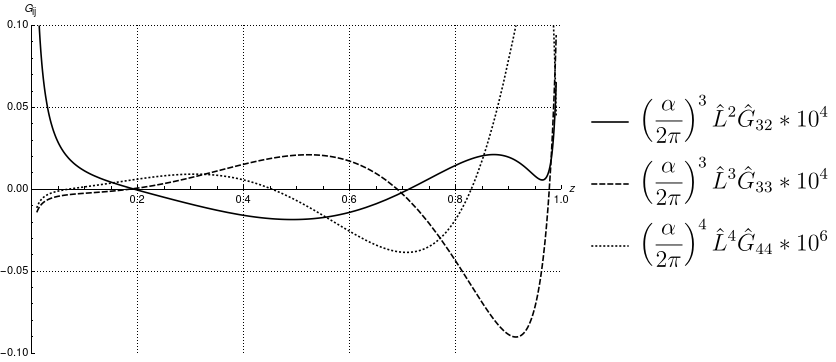}
		\caption{$G$-part of the corrections of the orders $\mathcal{O}(\alpha^3 L^2,\ \alpha^3 L^3,\ \alpha^4 L^4)$ for the new factorization scale.} \label{G3}
\end{figure}

We present our results for the new factorization scale for $F$ (Figs. \ref{F3}) 
and $G$ (Figs. \ref{G1}, \ref{G3}) functions. We take corrections of the orders $\mathcal{O}(\alpha)$ 
and $\mathcal{O}(\alpha L)$ from~\cite{Kinoshita:1958ru}, and contributions 
of the order $\mathcal{O}(\alpha)$ are recalculated with the new factorization scale.

\newpage



We also can look at the relative values of the contributions of different orders 
\begin{equation}
\left( \frac{\alpha}{2 \pi}\right)^i \hat{L}^j \frac{\hat{F}_{ij} (z)}{f_0 (z)}
\end{equation}
and the relative value of the full correction $F$ (in the unpolarized case) 
\begin{equation} \label{F_relative}
\delta_F=\frac{1}{f_0 (z)} \sum_{i,j}  \left( \frac{\alpha}{2 \pi}\right)^i \hat{L}^j \hat{F}_{ij} (z). 
\end{equation}
We can't show the same picture for $G$ part because of the zero of the function $g_0 (z)$ at $z = 0.5$, but the effect would be of the same order.

\begin{figure}[h!]
 \includegraphics[width=0.9\linewidth]{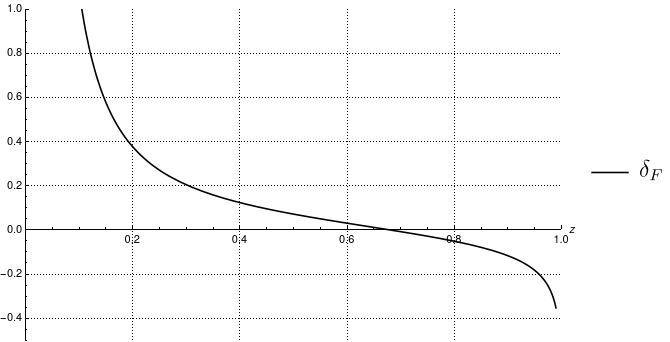}
		\caption{The full relative correction~(\ref{F_relative}) for the unpolarized case.} \label{RelFull}
\end{figure}

\begin{figure}[h!]
 \includegraphics[width=0.9\linewidth]{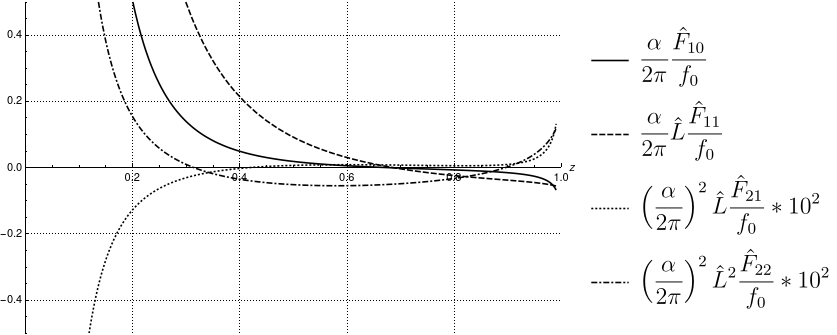}
		\caption{$F$-part contributions relative to $f_0$ in the first two orders.} \label{Rel12}
\end{figure}
\begin{figure}[h!]
		\includegraphics[width=0.9\linewidth]{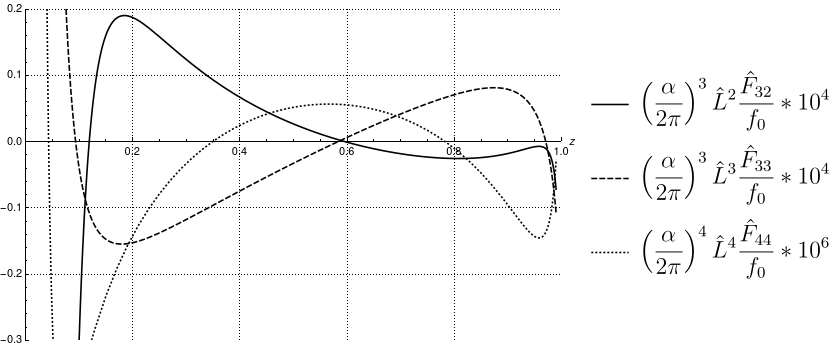}
		\caption{Higher-order $F$-part contributions relative to $f_0$.} \label{Rel34}
\end{figure}


\section{Conclusions}

In this way, we computed radiative corrections to polarized and non-polarized muon decay spectrum in 
the $\mathcal{O}(\alpha^3 L^2)$ and $\mathcal{O}(\alpha^4 L^4)$ orders. 
A new factorization scale is chosen to improve the convergence of the expansion in the powers 
of the large logarithm and thus suppress unknown NNLO effects. With this factorization scale we 
have the large logarithm $\hat{L}=\ln\frac{m^2_\mu z}{m_e^2(1-z)}$ where $z$ is the energy fraction of 
the electron in the final state.

Two mistakes in earlier results in the $\mathcal{O}(\alpha^2 L)$ and $\mathcal{O}(\alpha^3 L^3)$ orders 
are corrected. 
Our results are relevant for high-precision experiments on muon decays.
The results can be easily adapted for leptonic modes of tau lepton decays.

\begin{acknowledgments}
We are grateful to A.~Signer and Y.~Ulrich for cross-checks and useful discussions. 
A.A. thanks the Russian Science Foundation for the support (project No. 22-12-00021).
\end{acknowledgments}

\appendix

\section{Splitting and fragmentation functions} \label{appendix:Functions}

The next-to-leading order electron splitting function can be divided into four parts:
\begin{eqnarray} 
&& P_{ee}^{(1, \gamma)} =   2 x  \ln x  + \ln x \ln (1-x)   \left(  - 2 + \frac{4}{1-x} - 2 x \right)   \nonumber \\
&&     + \ln^2 x   \left( \frac{5}{2} - \frac{4}{1-x} + \frac{5}{2} x \right) + 2 - \frac{4}{1-x} \mathrm{Li}_2 (1-x)\nonumber \\
&&  + 2 \mathrm{Li}_2 (1-x) - 3 x + 2 x \mathrm{Li}_2 (1-x),    \\         
&& P_{ee}^{(1, NS)} = \ln x   \left( \frac{2}{3} - \frac{4}{3 (1-x)} + \frac{2}{3} x \right)  - \frac{4}{3} + \frac{4}{3} x,\\
&& P_{ee}^{(1, S)} =  \ln x   \left(  - 5 - 9 x - \frac{8}{3} x^2 \right)  + \ln^2 x   ( 1 + x )
       - 8 - \frac{20}{9 x} \nonumber \\
&& + 4 x + \frac{56}{9} x^2, \\
&& P_{ee}^{(1, int)} = \ln x   \left(  - 5 + \frac{3}{1-x} - 5 x \right) - 7 + \frac{4}{1-x} \mathrm{Li}_2 (1-x) \nonumber \\
&& - 2 \mathrm{Li}_2 (1-x) + 8 x - 2 x \mathrm{Li}_2 (1-x).
\end{eqnarray}

Note that we have removed the term $\frac{10}{9} P_{ij}^{(0)}$ from the expressions for functions $P_{ij}^{(1,NS)}$ given in refs.~\cite{Arbuzov:2002rp, Arbuzov:2002cn} because it naturally comes from the running coupling constant and can be kept there
as discussed in~\cite{Arbuzov:2022fmv}.

The fragmentation function $[D_{ee}]_T$ can also be divided into four parts:
\begin{equation} 
 \bigl[ D_{ee}\bigr]_T = \bigl[ D_{ee}^{ \gamma} \bigr]_T
\! + \!\bigl[ D_{ee}^{S} \bigr]_T  
\! + \! \bigl[ D_{ee}^{NS} \bigr]_T
\! + \! \bigl[ D_{ee}^{int} \bigr]_T
\! + \! \mathcal{O}( \alpha^4),
\end{equation}

\begin{eqnarray} 
&& \bigl[ D_{ee}^{ \gamma} \bigr]_T =  \delta (1-x)  + \frac{\alpha}{2 \pi} d_{ee}^{(1)}(x) + \frac{\alpha}{2 \pi}  L  P_{ee}^{(0)} 
\nonumber \\
&& \quad  + \left(\frac{\alpha}{2 \pi} \right)^2 L \biggl( P_{ee}^{(1, \gamma)}  	+ P_{ee}^{(0)} \otimes d_{ee}^{(1)} \biggr)    \nonumber \\
&& \quad     + \Bigl( \frac{\alpha}{2 \pi} \Bigr)^2 L^2 
	\Bigl( P_{ee}^{(0)} \otimes P_{ee}^{(0)}\Bigr) + \left(\frac{\alpha}{2 \pi} \right)^3 L^2 \biggl( \frac{2}{3} P_{ee}^{(1, \gamma)}  \nonumber \\
&& \quad + P_{ee}^{(0)} \otimes P_{ee}^{(1, \gamma)} + \frac{1}{3} P_{ee}^{(0)} \otimes d_{ee}^{(1)} + \frac{1}{2} P_{ee}^{(0)} \otimes P_{ee}^{(0)} \otimes d_{ee}^{(1)}\biggr)  \nonumber \\
&& \quad + \Bigl( \frac{\alpha}{2 \pi} \Bigr)^3 L^3  \frac{1}{6} P_{ee}^{(0)} \otimes P_{ee}^{(0)} \otimes P_{ee}^{(0)}, \\
&& \bigl[ D_{ee}^{NS} \bigr]_T =  \left(\frac{\alpha}{2 \pi} \right)^2 L \left(P_{ee}^{(1), NS} -  \frac{10}{9} P_{ee}^{(0)} \right) \nonumber \\
&& \quad + \left(\frac{\alpha}{2 \pi} \right)^2 L^2 \frac{1}{3} P_{ee}^{(0)}  + \left(\frac{\alpha}{2 \pi} \right)^3 L^2 \biggl( \frac{2}{3} P_{ee}^{(1, NS)}  \nonumber \\
&& \quad + P_{ee}^{(0)} \otimes P_{ee}^{(1, NS)} - \frac{13}{54} P_{ee}^{(0)} - \frac{10}{9} P_{ee}^{(0)} \otimes P_{ee}^{(0)}\biggr) \nonumber \\
&& \quad + \left(\frac{\alpha}{2 \pi} \right)^3 L^3 \left( \frac{1}{3} P_{ee}^{(0)} \otimes P_{ee}^{(0)} + \frac{4}{27} P_{ee}^{(0)} \right),\\
&& \bigl[ D_{ee}^{ S} \bigr]_T = \left(\frac{\alpha}{2 \pi} \right)^2 L \left(P_{ee}^{(1), S} +  d_{\gamma e}^{(1)}(x) \otimes P_{e \gamma}^{(0)} \right) \nonumber \\
&& + \Bigl( \frac{\alpha}{2 \pi} \Bigr)^2 L^2 
	 \frac{1}{2} P_{\gamma e}^{(0)} \otimes P_{e \gamma}^{(0)}  +   \Bigl( \frac{\alpha}{2 \pi} \Bigr)^3 L^2 \Bigl( \frac{1}{2} P_{e \gamma}^{(0)} \otimes P_{\gamma e}^{(1)} \nonumber \\
&& \quad + \frac{1}{2} P_{\bar{e} e}^{(0)} \otimes P_{e \bar{e}}^{(1)} + \frac{1}{3} d_{\gamma e}^{(1)} \otimes P_{e \gamma}^{(0)} + \frac{1}{2} d_{\gamma e}^{(1)} \otimes P_{\gamma \gamma}^{(0)} \otimes P_{e \gamma}^{(0)} \nonumber \\
&& \quad + \frac{1}{2} P_{\gamma e}^{(0)} \otimes P_{e \gamma}^{(1)}   - \frac{10}{9} P_{\gamma e}^{(0)} \otimes P_{e \gamma}^{(0)} +\frac{2}{3} P_{ee}^{(1, S)} \nonumber \\
&& \quad + \frac{1}{2} d_{ee}^{(1)} \otimes P_{\gamma e}^{(0)} \otimes P_{e \gamma}^{(0)}
 + \frac{1}{2} d_{\gamma e}^{(1)} \otimes P_{e e}^{(0)} \otimes P_{e \gamma}^{(0)} \nonumber \\
&& \quad +P_{ee}^{(0)} \otimes P_{ee}^{(1, S)} \Bigr) + \Bigl( \frac{\alpha}{2 \pi} \Bigr)^3 L^3 \biggl(\frac{1}{6}  P_{ee}^{(0)} \otimes P_{ee}^{(0)} \otimes P_{ee}^{(0)} \nonumber \\
&& \quad+ \frac{2}{9} P_{\gamma e}^{(0)} \otimes  P_{e \gamma}^{(0)} \biggr), \label{DeeS} \\
&& \bigl[ D_{ee}^{ int} \bigr]_T = \left(\frac{\alpha}{2 \pi} \right)^2 L P_{ee}^{(1, int)} + \left(\frac{\alpha}{2 \pi} \right)^3 L^2 \biggl( \frac{2}{3} P_{ee}^{(1, int)} \nonumber \\
&& \quad + P_{ee}^{(0)} \otimes P_{ee}^{(1, int)}\biggr).
\end{eqnarray}

The $\bigl[ D_{e \gamma} \bigr]_T$ fragmentation function reads
\begin{eqnarray}
&&	\bigl[ D_{e \gamma} \bigr]_T = \frac{\alpha}{2 \pi}  d_{e \gamma}^{(1)} + \frac{\alpha}{2 \pi} L  ( P_{e \gamma}^{(0)} ) + \left( \frac{\alpha}{2 \pi} \right)^2 L \Bigl( P_{e \gamma}^{(1) T} - \frac{10}{9} P_{e \gamma}^{(0)}  \nonumber \\
&&  \quad
+ P_{ee}^{(0)} \otimes d_{e \gamma}^{(1)} \Bigr) 	+ \left( \frac{\alpha}{2 \pi} \right)^2 L^2 \biggl( \frac{1}{3} P_{e \gamma}^{(0)} + \frac{1}{2} P_{\gamma \gamma}^{(0)} \otimes P_{e \gamma}^{(0)}  \nonumber \\
&&  \quad+ \frac{1}{2} P_{ee}^{(0)} \otimes P_{e \gamma}^{(0)} \biggr) + {\mathcal{O}}(\alpha^3).
\end{eqnarray}

\section{Analytic results} \label{results}

Here we present the analytic formulae for the computed higher-order contributions which appear in Eq.~(\ref{eq:F_z}).
\begin{eqnarray}
&& F_{21}(z) = -\frac{4405}{216} + \frac{2 \zeta_2 z^3}{3}-9 \zeta_2 z^2+\biggl(8 \zeta_2 z^3-12 \zeta_2 z^2-\frac{32 z^3}{9} \nonumber \\
&& -19 z^2-13 z-\frac{97}{12} \biggr) \ln (z) +12 \zeta_2 z+\left(8 z^3-12 z^2\right) \biggl(-\mathrm{Li}_3(z)\nonumber \\
&& +\mathrm{Li}_2(z) \ln (z) +\frac{1}{2} \ln (1-z) \ln ^2(z)+\zeta_3\biggr)+\biggl(-\frac{16 z^3}{3}+6 z^2\nonumber \\
&&-6 z\biggr) \mathrm{Li}_2(1-z)+\left(24 z^2-16 z^3\right) \mathrm{Li}_3(1-z) +\left(16 z^3-24 z^2\right)  \nonumber \\
&& \times \mathrm{Li}_2(1-z) \ln (1-z)+\left(8 z^3-12 z^2\right) \mathrm{Li}_2(1-z) \ln (z) \nonumber \\
&& -12 z^3 \zeta_3 -\frac{167 z^3}{54}+\left(\frac{16 z^3}{3}-12
   z\right) \ln ^2(1-z)+18 z^2 \zeta_3\nonumber \\
&&+\frac{449 z^2}{9}+\left(12 z^2-8 z^3\right) \ln ^3(z) +\biggl(-\frac{32 z^3}{3}+11 z^2-3 z\nonumber \\
&& -  \frac{5}{4}\biggr)
   \ln ^2(z)+\left(24 z^3-36 z^2\right) \ln (1-z) \ln ^2(z) +\left(12 z^2-8 z^3\right)  \nonumber \\
&& \times \ln ^2(1-z) \ln (z)+\left(-\frac{8 z^3}{9}+\frac{4
   z^2}{3}-16 z+\frac{2}{3 z}  -\frac{8}{3}\right) \ln (1-z) \nonumber \\
&&+\left(\frac{8 z^3}{3}-14 z^2+22 z+\frac{20}{3}\right) \ln (1-z) \ln (z)-\frac{1195 z}{36}-\frac{3}{z},
\end{eqnarray}

\begin{eqnarray}
&& G_{21}(z)=\left( \frac{2  z^3}{3}+12 z+\frac{8 }{3 z}-8 -11  z^2 \right) \zeta_2+\biggl(8 \zeta_2 z^3-4 \zeta_2 z^2\nonumber \\
&& \quad -\frac{44 z^3}{9}-\frac{56 z^2}{9}-\frac{26
   z}{3}+\frac{49}{12}\biggr) \ln (z) +\left(8 z^3-4 z^2\right)
   \biggl(-\mathrm{Li}_3(z)\nonumber \\
&& \quad+\mathrm{Li}_2(z) \ln (z)+\frac{1}{2} \ln (1-z) \ln ^2(z)+\zeta_3\biggr)+\biggl(-\frac{16 z^3}{3}+6 z^2 \nonumber \\
&& \quad -6 z -\frac{8}{3z}+\frac{13}{3}\biggr) \mathrm{Li}_2(1-z)+\left(8 z^2-16 z^3\right) \bigl( \mathrm{Li}_3(1-z) \nonumber \\
&& \quad- \mathrm{Li}_2(1-z) \ln (1-z) \bigr)  +\left(8
   z^3-4 z^2\right) \mathrm{Li}_2(1-z) \ln (z)-12 z^3 \zeta_3\nonumber \\
&& \quad-\frac{83 z^3}{18}+6 z^2 \zeta_3+\frac{1025 z^2}{54}+\left(4 z^2-8 z^3\right) \ln
   ^3(z) +\biggl(\frac{16 z^3}{3}+8 z^2 \nonumber \\
&& \quad -12 z-\frac{8}{3 z}+8\biggr) \ln ^2(1-z)+\biggl(-\frac{32 z^3}{3}+\frac{43 z^2}{3} +\frac{1}{4}\biggr) \ln^2(z)\nonumber \\
&& \quad+\left(24 z^3-12 z^2\right) \ln (1-z) \ln ^2(z)+\left(4 z^2-8 z^3\right) \ln ^2(1-z) \ln (z) \nonumber \\
&&\quad+\left(\frac{4 z^3}{9}+\frac{98
   z^2}{9}+\frac{28 z}{3}-\frac{10}{9 z}-2\right) \ln (1-z) +\biggl(\frac{8 z^3}{3}-\frac{86 z^2}{3}\nonumber \\
&& \quad +6 z-\frac{16}{3} \biggr) \ln (1-z) \ln
   (z)-\frac{137 z}{12}+\frac{29}{27 z}+\frac{415}{72}, 
\end{eqnarray}

\begin{eqnarray}
&& F_{32} = \frac{53623}{1296}+\frac{1}{108 z} +\frac{1201 z^3}{162}-\frac{2131 z^2}{72}-\frac{49 z}{2} \nonumber \\
&& \quad+\left(8 z^3-4 z^2-12 z\right) \ln ^3(1-z) +\biggl(\frac{92 z^3}{9}-\frac{41 z^2}{3}-\frac{7
   z}{3} \nonumber \\
&& \quad-\frac{35}{36}\biggr) \ln ^3(z)+\biggl[\frac{142 z^3}{9}+\frac{152 z^2}{3}+\frac{161 z}{12}+\left(4 z^3-6 z^2\right) \nonumber \\
&& \quad\times \ln ^2(1-z)+\zeta_2
   \left(60 z^2-40 z^3\right)+\biggl(-\frac{56 z^3}{3}+58 z^2+44 z \nonumber \\
&& \quad+\frac{125}{6}\biggr) \ln (1-z)+\frac{37}{8} \biggr] \ln ^2(z)+\left(4 z^3-6
   z^2\right) \mathrm{Li}_2(1-z){}^2 \nonumber \\
&& \quad+\zeta_2 \left(-6 z^2+16 z+\frac{139}{18}\right) + (2z^3 - 3z^2)\biggl\{- 20 \zeta_4 
\nonumber \\
&& \quad- 18 \zeta_2^2 + 52 \left( \text{H}(3,0,z) + \text{H}(2,0,0,z) + \text{H}(2,1,0,z)\right)\nonumber \\
&& \quad+ 40  \ ( \text{H}(1,2,0,z) + \text{H}(1,1,1,0,z)) + 32 \ \text{H}(0,0,0,0,z)
\nonumber \\
&& \quad
   + 28 \ \text{H}(1,0,0,0,z)+48 \ \text{H}(1,1,0,0,z)  \biggr \} +\biggl(\frac{136 z^3}{9}\nonumber \\
&& \quad+\frac{185 z^2}{3}-\frac{247 z}{3} +\zeta_2 \left(60 z^2-40
   z^3\right)-\frac{160}{3} -\frac{6}{z}\biggr)\nonumber \\
&& \quad \times \mathrm{Li}_2(1-z)  +\ln ^2(1-z)  \biggl(-\frac{62 z^3}{9}+\frac{37 z^2} {6} -\frac{62 z}{3} \nonumber \\
&& \quad +\zeta_2
   \left(40 z^3-60 z^2\right)  +\left(16 z^3-24 z^2\right) \mathrm{Li}_2(1-z)  \nonumber \\
&& \quad -\frac{121}{18}+\frac{1}{z}\biggr) +\biggl(32 z^3-40 z^2   +4 z-\frac{10}{3}\biggr)
   \mathrm{Li}_3(1-z)\nonumber \\
&& \quad  +\left(\frac{64 z^3}{3}+72 z^2+97 z+\frac{140}{3}\right) \mathrm{Li}_3(z)  \nonumber \\
&& \quad  +\left(72 z^2-48 z^3\right)\mathrm{Li}_4(1-z)+\left(120
   z^3-180 z^2\right) \mathrm{Li}_4(z)  \nonumber \\
&& \quad +\left(120 z^2-80 z^3\right) \mathrm{S}_{2,2}(z) +\ln (z) \biggl[ \frac{283 z^3}{27}-\frac{799 z^2}{12} \nonumber \\
&& \quad+\frac{539
   z}{36}+\left(12 z^2-8 z^3\right) \ln ^3(1-z) +\biggl(-\frac{16 z^3}{3} -8 z^2 \nonumber \\
&& \quad+36 z+10\biggr) \ln ^2(1-z)  +\zeta_2 \biggl(-\frac{100 z^3}{3} -32 z^2 \nonumber \\
&&\quad-125 z-\frac{160}{3}\biggr)+\left(\frac{8 z^3}{3}+22 z^2+74 z+\frac{185}{6}\right) \mathrm{Li}_2(1-z) \nonumber \\
&&\quad+\ln (1-z) \biggl(-\frac{4 z^3}{3}-11
   z^2-\frac{z}{2}+\zeta_2 \left(16 z^3-24 z^2\right) \nonumber \\
&& \quad+\left(8 z^3-12 z^2\right) \mathrm{Li}_2(1-z)-\frac{57}{4}+\frac{8}{3 z}\biggr) +\left(48 z^2-32
   z^3\right) \zeta_3\nonumber \\
&& \quad+ \frac{1261}{108}\biggr]+\ln (1-z) \biggl[ -\frac{155 z^3}{27}+\frac{2221 z^2}{36}-\frac{677 z}{36} \nonumber \\
&& \quad+\zeta_2 \biggl(-\frac{20
   z^3}{3}-14 z^2+36 z\biggr) +\left(-32 z^3+40 z^2-4 z+\frac{10}{3}\right) 
\nonumber \\
&&\quad \times \mathrm{Li}_2(1-z) +\left(48 z^2-32 z^3\right) \mathrm{Li}_3(1-z)+\left(144
   z^2-96 z^3\right)  \nonumber \\
&& \quad \times \mathrm{Li}_3(z) +\left(88 z^3-132 z^2\right) \zeta_3-\frac{6281}{108}-\frac{32}{9 z} \biggr]+\biggl(\frac{8 z^3}{3}-92 z^2\nonumber \\
&& \quad -109
   z-\frac{125}{3}\biggr) \zeta_3,
\end{eqnarray}

\begin{eqnarray}
&& G_{32}(z) = \frac{11329}{1296}-\frac{131}{108 z} -\frac{55 z^2}{24}-\frac{421 z}{54}+ \frac{1273 z^3}{162}\nonumber \\
&& \quad +(2 z-1)z^2 \left( 52 \text{H}(3,0,z) +40 \text{H}(1,2,0,z) \right) +\biggl(8
   z^3+\frac{20 z^2}{3} \nonumber \\
&& \quad  -12 z+8-\frac{8}{3 z}\biggr) \ln ^3(1-z)+\left(\frac{92 z^3}{9}-\frac{89 z^2}{9}+\frac{7}{36}\right) \ln
   ^3(z)\nonumber \\
&&\quad +\biggl(\frac{148 z^3}{9}+11 z^2+\frac{49 z}{12} +\zeta_2 \left(20 z^2-40 z^3\right) \nonumber \\
&& \quad +\left(-\frac{64 z^3}{3}+48 z^2-6 z+6-\frac{4}{3
   z}\right) \ln (1-z)-\frac{65}{24}\biggr) \ln ^2(z) \nonumber \\
&&\quad  +\left(4 z^3-2 z^2\right) \mathrm{Li}_2(z){}^2 +\zeta_2 \biggl(-\frac{122 z^3}{9}+\frac{40
   z^2}{9}-30 z+\frac{115}{18}\nonumber \\
&& \quad +\frac{26}{9 z}\biggr) +\zeta_4 \left(40 z^3-20 z^2\right)+\left(104 z^3-52 z^2\right)
   \text{H}(2,0,0,z)\nonumber \\
&& \quad  +\left(104 z^3-52 z^2\right) \text{H}(2,1,0,z)+\left(64 z^3 \! - \! 32 z^2\right) \text{H}(0,0,0,0,z)\nonumber \\
&& \quad  +\left(56 z^3 \!- \!28
   z^2\right) \text{H}(1,0,0,0,z) +\left(96 z^3 \! - \!48 z^2\right) \text{H}(1,1,0,0,z)\nonumber \\
&& \quad +\left(80 z^3-40 z^2\right)
   \text{H}(1,1,1,0,z)+\biggl(26 z^3-13 z^2 \nonumber \\
&& \quad 
   +\zeta_2 \left(16 z^2-32 z^3\right)\biggr) \mathrm{Li}_2(1-z)+\biggl(\frac{110 z^3}{9}-\frac{112
   z^2}{9}+16 z\nonumber \\
&& \quad -\frac{5}{3}-\frac{2}{z}\biggr) \mathrm{Li}_2(z)+\ln ^2(1-z) \biggl[-\frac{50 z^3}{9} +\frac{401 z^2}{18} +16 z \nonumber \\
&& \quad +\left(56 z^3-28 z^2\right)
   \mathrm{Li}_2(1-z)+\left(40 z^3-20 z^2\right) \mathrm{Li}_2(z) \nonumber \\
&& \quad -\frac{79}{18}-\frac{11}{9 z}\biggr]+\biggl(32 z^3-16 z^2 +12
   z-\frac{22}{3}+\frac{8}{z}\biggr) \mathrm{Li}_3(1-z) \nonumber \\
&& \quad +\left(\frac{64 z^3}{3}+\frac{56 z^2}{3}-15 z+\frac{2}{3}+\frac{8}{3 z}\right)   \mathrm{Li}_3(z) \nonumber  \\
&& \quad +\left(24 z^2-48 z^3\right) \mathrm{Li}_4(1-z)  +\left(120 z^3-60 z^2\right) \mathrm{Li}_4(z)\nonumber \\
&& \quad +\left(40 z^2-80 z^3\right) \mathrm{S}_{2,2}(z)+\ln
   (z) \biggl[\frac{319 z^3}{27}-\frac{4205 z^2}{108}-\frac{25 z}{12} \nonumber \\
&& \quad + \left(32 z^3-16 z^2\right) \ln ^3(1-z)+\left(-\frac{16 z^3}{3}-40 z^2+12
   z-10\right)\nonumber \\
&& \quad \times \ln ^2(1-z)+\zeta_2 \left(-\frac{92 z^3}{3}+\frac{94 z^2}{3}+3 z+\frac{5}{2}\right) +\biggl(\frac{74 z^3}{9} \nonumber \\
&& \quad -\frac{253
   z^2}{9}-\frac{41 z}{6}+\zeta_2 \left(24 z^3-12 z^2\right)+\frac{5}{4}-\frac{2}{z}\biggr) \ln (1-z)
\nonumber \\
&& \quad  
   +\left(-\frac{8 z^3}{3}+\frac{2
   z^2}{3}+\frac{37}{6}-\frac{8}{3 z}\right) \mathrm{Li}_2(z) +\left(16 z^2-32 z^3\right) \zeta_3\nonumber \\
&& \quad +\frac{136}{27}\biggr] +\ln (1-z) \biggl[ -\frac{191
   z^3}{27}+\frac{2741 z^2}{108}+\frac{19 z}{12}+\zeta_2 \biggl(-\frac{20 z^3}{3} \nonumber \\
&& \quad - \frac{86 z^2}{3}+36 z-24+\frac{8}{z}\biggr) -\left(32 z^3-16
   z^2+12 z-\frac{22}{3}+\frac{8}{z}\right)  \nonumber \\
&& \quad \times \mathrm{Li}_2(1-z)
   +\left(16 z^2-32 z^3\right) \mathrm{Li}_3(1-z)+(48 z^2 \nonumber \\
&& \quad -96 z^3)
   \mathrm{Li}_3(z) +\left(88 z^3-44 z^2\right) \zeta_3+\frac{1}{108}+\frac{14}{3 z}\biggr]+\biggl(\frac{8 z^3}{3} \nonumber \\
&& \quad  -\frac{76 z^2}{3}-9
   z  +\frac{43}{3}-\frac{8}{z}\biggr) \zeta_3,
\end{eqnarray}

\begin{eqnarray}
&& F_{33} (z) = -\frac{619}{1296} +\frac{20 z}{9}+\frac{4}{27 z}+ \ln (z) \biggl[\zeta_2 \left(12 z^2-8 z^3\right)\nonumber \\
&& \quad+\left(\frac{16 z^3}{3}-8 z^2\right) \mathrm{Li}_2(z)+\frac{32 z^3}{27}+\frac{52 z^2}{9} +\left(8
   z^3-12 z^2\right)\nonumber \\
&& \quad \times \ln ^2(1-z)+\left(-\frac{8 z^3}{3}+4 z^2-4 z-\frac{5}{3}\right) \ln (1-z)\nonumber \\
&& \quad -\frac{67 z}{36}-\frac{41}{108}\biggr]  +\ln (1-z)
   \biggl[\zeta_2 \left(8 z^3-12 z^2\right)  \nonumber \\
&& \quad +\left(8 z^3-12 z^2\right) \mathrm{Li}_2(1-z)-\frac{32 z^3}{27}-\frac{44 z^2}{9}+\frac{15 z}{2}+\frac{4}{9
   z} \nonumber \\
&& \quad +\frac{289}{108}\biggr] +\zeta_2 \left(-\frac{16 z^3}{3}+\frac{8 z^2}{3}+\frac{2 z}{3}+\frac{5}{18}\right)+\biggl(\frac{8 z^3}{3}+\frac{4
   z^2}{3}\nonumber \\
&& \quad -\frac{14 z}{3}-\frac{35}{18}\biggr) \mathrm{Li}_2(z)+\left(12 z^2-8 z^3\right) \mathrm{Li}_3(1-z) \nonumber \\
&& \quad +\left(8 z^2-\frac{16 z^3}{3}\right)
   \mathrm{Li}_3(z)+\frac{16 z^3}{81}-\frac{5 z^2}{9}+\left(4 z^2-\frac{8 z^3}{3}\right) \nonumber \\
&& \quad\times \ln ^3(1-z) +\left(\frac{4 z^3}{9}-\frac{2 z^2}{3}\right) \ln
   ^3(z) +\biggl(\frac{8 z^3}{3}-4 z^2+4 z \nonumber \\
&& \quad +\frac{5}{3}\biggr) \ln ^2(1-z)+\biggl(\frac{4 z^3}{3}-\frac{2 z^2}{3}+\biggl(2 z^2 -\frac{4 z^3}{3}\biggr) \ln
   (1-z) \nonumber \\
&& \quad  -\frac{z}{2}-\frac{5}{24}\biggr) \ln ^2(z),
\end{eqnarray}

\begin{eqnarray}
&& G_{33} (z) = -\frac{49}{1296}-\frac{115
   z}{108}-\frac{4}{81 z} +\ln (z) \biggl[ \zeta_2 \left(\frac{4 z^2}{3}-\frac{8 z^3}{3}\right)\nonumber \\
&& \quad +\left(\frac{8 z^2}{3}-\frac{16 z^3}{3}\right) \mathrm{Li}_2(1-z)+\frac{32
   z^3}{27} +\frac{68 z^2}{27} +\left(8 z^3-4 z^2\right) \nonumber \\
&& \quad   \times \ln ^2(1-z)+\left(-\frac{16 z^3}{3}+\frac{56 z^2}{9}-\frac{1}{18}\right) \ln
   (1-z) \nonumber \\
&& \quad +\frac{z}{12}+\frac{1}{108} \biggr] +\ln (1-z) \biggl(\zeta_2 \left(8 z^3-4 z^2\right)  +\biggl(8 z^3-4 z^2\biggr)\nonumber \\
&&  \quad \times \mathrm{Li}_2(1-z)-\frac{32
   z^3}{27}  -\frac{20 z^2}{9}-\frac{z}{2}-\frac{4}{27 z}-\frac{53}{108}\biggr) \nonumber \\
&&  \quad +\zeta_2 \left(-\frac{8 z^3}{3}+\frac{20
   z^2}{3}+\frac{1}{3}\right)+\left(-\frac{8 z^3}{3}-\frac{4 z^2}{9}-\frac{7}{18}\right) \nonumber \\
&& \quad \times \mathrm{Li}_2(1-z) +\left(4 z^2-8 z^3\right)
   \mathrm{Li}_3(1-z) +\left(\frac{8 z^2}{3}-\frac{16 z^3}{3}\right) \nonumber \\
&& \quad \times \mathrm{Li}_3(z)+\frac{16 z^3}{81}-\frac{47 z^2}{81}+\left(\frac{4 z^2}{3}-\frac{8
   z^3}{3}\right) \ln ^3(1-z)  \nonumber \\
&& \quad+ \left(\frac{4 z^3}{9}-\frac{2 z^2}{9}\right) \ln ^3(z)  +\left(\frac{8 z^3}{3}-\frac{20 z^2}{3}-\frac{1}{3}\right) \ln
   ^2(1-z) \nonumber \\
&& \quad +\biggl(\! \frac{4 z^3}{3} \! - \! \frac{10 z^2}{9} \! + \!\left(\! \frac{10 z^2}{3} \! - \! \frac{20 z^3}{3} \!\right) \ln (1 \!-\! z)+\frac{1}{24} \! \biggr) \ln ^2(z),
\end{eqnarray}

\begin{eqnarray}
&& F_{44}(z) = -\frac{7577}{3456}+\left(2 z^2-\frac{4 z^3}{3}\right) \ln ^4(1-z)+\biggl(\frac{8 z^3}{9}-\frac{4 z^2}{3}\nonumber \\
&& \quad+\frac{8 z}{3}+\frac{10}{9}\biggr) \ln ^3(1-z)+\biggl (\zeta_2 \left(8 z^3-12 z^2\right)-\frac{8
   z^3}{27}-\frac{47 z^2}{9}  \nonumber \\
&& \quad +\frac{25 z}{3} +\left(12 z^2-8 z^3\right) \mathrm{Li}_2(z)+\frac{35}{12}+\frac{1}{3
   z}\biggr) \ln ^2(1-z)  \nonumber \\
&& \quad+\biggl(-\frac{157 z^2}{54}+\frac{725 z}{108}  +\zeta_2 \left(-\frac{8 z^3}{3}+4 z^2-8 z-\frac{10}{3}\right) \nonumber \\
&& \quad +\left(-\frac{8
   z^3}{3}-8 z^2+8 z+\frac{10}{3}\right) \mathrm{Li}_2(1-z)+\left(24 z^2-16 z^3\right)  \nonumber \\
&& \quad \times \mathrm{Li}_3(1-z) +\left(\frac{16 z^3}{3}-8 z^2\right)
   \mathrm{Li}_3(z)+\left(32 z^2-\frac{64 z^3}{3}\right) \zeta_3 
 \nonumber \\
&& \quad +\frac{41}{72}+\frac{10}{27 z}\biggr) \ln (1-z)+\biggl(-\frac{4
   z^3}{27}-\frac{z^2}{9}-\frac{z}{24} -\frac{5}{288}\biggr) \ln ^3(z)\nonumber \\
&& \quad +\frac{278 z^2}{81}+\biggl(-\frac{4 z^3}{27}-\frac{43 z^2}{18}+\frac{5
   z}{96}+\zeta_2 \left(\frac{8 z^3}{3}-4 z^2\right) \nonumber \\
&& \quad
 +\left(\frac{4 z^3}{9}-\frac{2 z^2}{3}+\frac{4 z}{3}+\frac{5}{9}\right) \ln   (1-z) -\frac{607}{1728}\biggr) \ln ^2(z) \nonumber \\
&& \quad +\left(8 z^2-\frac{16 z^3}{3}\right) \mathrm{Li}_2(z){}^2+\frac{191 z}{576}+\zeta_2 \biggl(\frac{16 z^3}{27}+10 z^2 -\frac{34 z}{9}\nonumber \\
&& \quad -\frac{73}{108}\biggr) + \zeta_4 \left(70
   z^2-\frac{140 z^3}{3}\right)  +4 (3-2 z) z^2 \text{H}(3,0,z) \nonumber \\
&& \quad+4 (3-2 z) z^2 \text{H}(1,2,0,z) +\left(8 z^2-\frac{16 z^3}{3}\right) \text{H}(2,0,0,z)
\nonumber \\
&& \quad
+\left(24 z^2-16 z^3\right)
   \text{H}(2,1,0,z)+\left(2 z^2-\frac{4 z^3}{3}\right) \text{H}(0,0,0,0z) \nonumber \\
&& \quad+\left(2 z^2-\frac{4 z^3}{3}\right)
   \text{H}(1,0,0,0z)+\left(8 z^2-\frac{16 z^3}{3}\right) \text{H}(1,1,0,0z) \nonumber \\
&& \quad +\left(24 z^2-16 z^3\right)
   \text{H}(1,1,1,0z) +\biggl(-\frac{8 z^3}{27}-\frac{43 z^2}{9}-\frac{41 z}{9} \nonumber \\
&& \quad +\zeta_2 \left(\frac{32 z^3}{3}-16
   z^2\right)-\frac{121}{54}-\frac{1}{3 z}\biggr) \mathrm{Li}_2(z)+\biggl(\frac{8 z^3}{3}+8 z^2 \nonumber \\
&& \quad -8 z-\frac{10}{3}\biggr) \mathrm{Li}_3(1-z)  +\biggl(\frac{16
   z^3}{9}+\frac{8 z^2}{3} -\frac{4 z}{3}-\frac{5}{9}\biggr) \mathrm{Li}_3(z) \nonumber \\
&& \quad+\left(32 z^3-48 z^2\right) \mathrm{Li}_4(1-z)+\left(8 z^2-\frac{16
   z^3}{3}\right) \mathrm{Li}_4(z) \nonumber \\
&& \quad +\left(16 z^3-24 z^2\right) \mathrm{S}_{2,2}(z)+\ln (z) \biggl[\left(4 z^2-\frac{8 z^3}{3}\right) \ln ^3(1-z) \nonumber \\
&& \quad +\left(-\frac{8
   z^3}{3}-2 z^2\right) \ln ^2(1-z) +\biggl(\frac{8 z^3}{27}+\frac{47 z^2}{9}-\frac{25 z}{3} \nonumber \\
&& \quad+\zeta_2 \left(8 z^2-\frac{16
   z^3}{3}\right)-\frac{35}{12}-\frac{1}{3 z}\biggr) \ln (1-z)+\frac{23 z^2}{18}-\frac{6415 z}{864} \nonumber \\
&& \quad +\zeta_2 \left(\frac{8 z^3}{3}+\frac{8
   z^2}{3}-\frac{4 z}{3}-\frac{5}{9}\right)+\biggl(-\frac{16 z^3}{9}-\frac{10 z^2}{3}+\frac{8 z}{3} \nonumber \\
&& \quad +\frac{10}{9}\biggr) \mathrm{Li}_2(z)-\frac{2}{27
   z}+\left(\frac{16 z^3}{3}-8 z^2\right) \zeta_3-\frac{997}{576}\biggr] \nonumber \\
&& \quad -\frac{37}{162 z}+\left(-\frac{16 z^2}{3}+\frac{20 z}{3}+\frac{25}{9}\right)
   \zeta_3,
\end{eqnarray}

\begin{eqnarray}
&& G_{44}(z) = \frac{1015}{10368} + \left(\frac{2 z^2}{3}-\frac{4 z^3}{3}\right) \ln ^4(1-z)+\biggl(\frac{8 z^3}{9} \nonumber \\
&& \quad -4 z^2-\frac{2}{9}\biggr) \ln ^3(1-z)+\biggl(-\frac{8
   z^3}{27}-\frac{119 z^2}{27}-\frac{2 z}{3} 
   \nonumber \\
&& \quad +\zeta_2 \left(8 z^3-4 z^2\right)+\left(4 z^2-8 z^3\right) \mathrm{Li}_2(z)-\frac{7}{12}-\frac{1}{9
   z}\biggr)  \nonumber \\
&& \quad \times \ln ^2(1-z)+\biggl[\zeta_2 \left(-\frac{8 z^3}{3}+12 z^2+\frac{2}{3}\right)-\frac{349 z^2}{162}   \nonumber \\
&& \quad -\frac{77 z}{36}+\left(-\frac{8
   z^3}{3}-\frac{8 z^2}{3}-\frac{2}{3}\right) \mathrm{Li}_2(1-z)+\left(8 z^2-16 z^3\right)   \nonumber \\
&& \quad\times \mathrm{Li}_3(1-z)+\left(\frac{16 z^3}{3}-\frac{8
   z^2}{3}\right) \mathrm{Li}_3(z)  +\left(\frac{32 z^2}{3}-\frac{64 z^3}{3}\right) \zeta_3  \nonumber \\
&& \quad-\frac{71}{216}-\frac{10}{81 z}\biggr] \ln
   (1-z) +\left(-\frac{4 z^3}{27}+\frac{z^2}{9}+\frac{1}{288}\right) \ln ^3(z)    \nonumber \\
&& \quad +\frac{26 z^2}{81}+\biggl(-\frac{4 z^3}{27}-\frac{25 z^2}{18}-\frac{41
   z}{288}+\zeta_2 \left(\frac{8 z^3}{3}-\frac{4 z^2}{3}\right)  \nonumber \\
&& \quad +\left(\frac{4 z^3}{9}-2 z^2-\frac{1}{9}\right) \ln (1-z)+\frac{107}{1728}\biggr)
   \ln ^2(z)    \nonumber \\
&&\quad +\left(\frac{8 z^2}{3}-\frac{16 z^3}{3}\right) \mathrm{Li}_2(z){}^2-\frac{9485 z}{5184}+\zeta_4 \left(\frac{70 z^2}{3}-\frac{140
   z^3}{3}\right)  \nonumber \\
&& \quad +\zeta_2 \left(\frac{16 z^3}{27}+\frac{62 z^2}{9}+\frac{z}{3}+\frac{5}{108}\right) +4 (1-2 z) z^2 \bigl( \text{H}(3,0,z) \nonumber \\
&& \quad  + \text{H}(1,2,0,z) \bigr)   + \left(\frac{8 z^2}{3}-\frac{16 z^3}{3}\right) \text{H}(2,0,0,z)   \nonumber \\
&& \quad +\left(8 z^2-16 z^3\right)
   \text{H}(2,1,0,z)+\left(\frac{2 z^2}{3}-\frac{4 z^3}{3}\right) \text{H}(0,0,0,0,z)   \nonumber \\
&& \quad +\left(\frac{2 z^2}{3}-\frac{4 z^3}{3}\right)
   \text{H}(1,0,0,0z)  +\left(\frac{8 z^2}{3}-\frac{16 z^3}{3}\right) \text{H}(1,1,0,0z)
\nonumber \\
&& \quad    
   +\left(8 z^2-16 z^3\right)
   \text{H}(1,1,1,0z)+\biggl(-\frac{8 z^3}{27}-\frac{67 z^2}{27}+\frac{z}{3}    \nonumber \\
&& \quad +\zeta_2 \left(\frac{32 z^3}{3}-\frac{16
   z^2}{3}\right)+\frac{29}{54}+\frac{1}{9 z}\biggr) \mathrm{Li}_2(z)+\biggl(\frac{8 z^3}{3}+\frac{8 z^2}{3}\nonumber \\
&& \quad +\frac{2}{3}\biggr)
   \mathrm{Li}_3(1-z)+\left(\frac{16 z^3}{9}-\frac{8 z^2}{3}+\frac{1}{9}\right) \mathrm{Li}_3(z)    \nonumber \\
&& \quad+\left(32 z^3-16 z^2\right) \mathrm{Li}_4(1-z)+\left(\frac{8
   z^2}{3}-\frac{16 z^3}{3}\right) \mathrm{Li}_4(z)  \nonumber \\
&& \quad  +\left(16 z^3-8 z^2\right) \mathrm{S}_{2,2}(z)   +\ln (z) \biggl[ \left(\frac{4 z^2}{3}-\frac{8 z^3}{3}\right)
   \ln ^3(1-z) \nonumber \\
&& \quad +\left(\frac{14 z^2}{3}-\frac{8 z^3}{3}\right) \ln ^2(1-z)+\biggl(\frac{8 z^3}{27}+\frac{119 z^2}{27}+\frac{2 z}{3}   +\frac{2}{81
   z} \nonumber \\
&& \quad
   +\zeta_2
   \left(\frac{8 z^2}{3}-\frac{16 z^3}{3}\right)+\frac{7}{12}+\frac{1}{9 z}\biggr) \ln (1-z)+\frac{71 z^2}{54}+\frac{19 z}{32}  \nonumber \\
&& \quad +\zeta_2
   \left(\frac{8 z^3}{3}-\frac{8 z^2}{3}+\frac{1}{9}\right)    +\left(-\frac{16 z^3}{9}+\frac{2 z^2}{3}-\frac{2}{9}\right) \mathrm{Li}_2(z) \nonumber \\
&& \quad+\left(\frac{16 z^3}{3}-\frac{8 z^2}{3}\right) \zeta_3+\frac{475}{1728}\biggr]+\frac{11}{162 z}-\left(\frac{16 z^2}{3}+\frac{5}{9}\right)
   \zeta_3.
\end{eqnarray}



\section{Harmonic polylogarithms}

Functions $H (a_1, ..., a_k;z)$ are harmonic polylogarithms \cite{Maitre:2005uu, Ablinger:2018sat}. 
The dimension of the vector $a = (a_1, ..., a_k)$ is called the weight of the harmonic polylogarithm (HPL). They are defined through the functions 
\begin{eqnarray}
\label{fk1}
 f_1 (z) = \frac{1}{1-z}, \\
 f_0 (z) = \frac{1}{z}, \\
 \label{fk3}
 f_{-1} (z) = \frac{1}{1+z},
\end{eqnarray}
We can get the HPLs recursively through the integration of the functions \\
(\ref{fk1})--(\ref{fk3}):
\begin{eqnarray}
&& H (1;z) = \int \limits_0^z f_1 (y) d y = - \ln (1-z), \\
&& H (0;z) = \ln z, \\
&& H (-1;z) = \int \limits_0^z f_{-1} (y) d y = - \ln (1+z), \\
&& H({}^n 0; z) = \frac{1}{n!} \ln^n z, \\
&& H (a,a_1, ..., a_k;z) = \int \limits_0^z f_a (y) H (a_1, ..., a_k;y) dy,
\end{eqnarray}
where 
\begin{equation}
{}^n 0 = \underbrace{0,...,0}_{n}.
\end{equation}

\bibliography{Muon-FG}

\end{document}